\documentclass[preprint]{aastex}
\usepackage{apjfonts,psfig}

\newcommand{\axaf}{\mbox{\it Chandra\/}}
\newcommand{\rxte}{\mbox{\it RXTE\/}}
\newcommand{\ergsec}{\mbox{erg s$^{-1}$}}

\slugcomment{To appear in the Astronomical Journal}
\shorttitle{Chandra Spectroscopy of the Rapid Burster}
\shortauthors{Marshall et al.}

\begin{document}

\title{The X-ray Spectrum of the Rapid Burster using
the {\em Chandra} HETGS}

\author{H.L. Marshall\altaffilmark{1},
R. Rutledge\altaffilmark{2},
D.W. Fox\altaffilmark{1},
J.M. Miller\altaffilmark{1},
R. Guerriero\altaffilmark{3},
E. Morgan\altaffilmark{1},
M. van der Klis\altaffilmark{4},
L.~Bildsten\altaffilmark{5},
T. Dotani\altaffilmark{6},
W.H.G. Lewin\altaffilmark{1} }
\altaffiltext{1}{Center for Space Research, Massachusetts Institute of
        Technology, Cambridge, MA 02139}
\altaffiltext{2}{Space Radiation Laboratory, Caltech, MS 220-47,
	Pasadena, CA 91125}
\altaffiltext{3}{Department of Physics,
	United States Military Academy, West Point, NY 10996}
\altaffiltext{4}{Faculty of Mathematics, Computer Science,
	Physics \& Astronomy, University of Amsterdam, the Netherlands}
\altaffiltext{5}{Harvard University, Cambridge, MA 02138}
\altaffiltext{6}{Institute of Space and Astronautical Science,
	Sagamihara, Kanagawa, 229-8510 Japan}
\email{hermanm@space.mit.edu}

\begin{abstract}

We present
observations of the Rapid Burster (RB, also known as
MXB 1730$-$335) using the
Chandra High Energy Transmission
Grating Spectrometer.  The average interval between
type II (accretion) bursts was about 40 s.
There was one type I (thermonuclear
flash) burst and about 20 ``mini-bursts'' which are
probably type II bursts whose
peak flux is 10-40\% of the average peak flux of the
other type II bursts.  The time averaged spectra of
the type II bursts are well fit by a blackbody with
a temperature of $kT = 1.6$ keV, a radius of
8.9 km for a distance of 8.6 kpc, and
an interstellar column density of
$1.7 \times 10^{22}$ cm$^{-2}$.
No narrow emission or absorption lines were clearly detected.
The 3 $\sigma$ upper limits to
the equivalent widths of any features are
$<$ 10 eV in the 1.1-7.0 keV band and as small as
1.5 eV near 1.7 keV.
We suggest that Comptonization destroys absorption
features such as the resonance line of Fe {\sc XXVI}.

\end{abstract}


\section{Introduction}

The Rapid Burster (RB) is a recurrent transient low-mass X-ray binary
with an X--ray burst phenomenology that makes it unique in the
Galaxy. The many idiosyncrasies of the RB are
overwhelming, and it is fair to say that we do not understand why the
RB, and only the RB, behaves so differently from all other low
mass X-ray binaries (LMXBs).
For a detailed review see \citet{lvt93}, hereafter LVT.  
The RB (MXB 1730$-$335) 
is the only low-mass X-ray
binary (LMXB) known to produce two types of X-ray bursts \citep{hml78}.
Type I bursts, which are observed from
$\sim$40 LMXBs, are due to thermonuclear flashes on the surface of an
accreting neutron star.  They typically recur on time scales of hours
to tens of hours (LVT).  Type II bursts 
are due to accretion
instabilities (gravitational potential energy).  In the RB, their
durations range from $\sim$2 to $\sim$680 sec, and thousands of short
bursts can occur in a single day.  The behavior of the type II bursts
from the RB is like that of a relaxation oscillator: the type II burst
fluence $E$ is roughly proportional to the time interval, $\Delta t$,
to the following burst (the ``$E$-$\Delta t$'' relation).  The time
averaged luminosity of the RB is $\sim$10$^{37}$ \ergsec\ during its
active periods which typically last for several weeks.  The type II
burst luminosities at burst maximum range from
$\sim$5$\times$10$^{37}$ to $\sim$4$\times$10$^{38}$ \ergsec\
(LVT).  Henceforward, whenever we use the word ``burst(s)'', we mean
type II burst(s) unless we specify otherwise.

Among the LMXBs, the RB stands alone in its complex QPO
behavior, none of which is understood (LVT).  
With the \textit{Rossi X-ray Timing Explorer\/} (\rxte) we may have
detected ($>98$\% significance) the signature of
near coherent 306.5 Hz pulsations in the average power-density
spectrum of the first second
of 31 Type I bursts \citep{fox2001}. 

Though we are largely at a loss in our understanding of this system,
it is important to remember that there have been occasions that,
when the RB became active, it behaved like an ordinary LMXB
and then transitioned to rapid bursting.
Starting August 5, 1983, no type II bursts were detected
but instead a strong persistent X-ray flux and type I bursts
were observed for several days \cite{kunieda84,barr87}.
In the period November 6--17, 1996, and July 7--10, 1997
(during a total of 18 ksec of \rxte\ observations) the RB again
produced only persistent emission and type I bursts at
the beginning of its activity cycle.

We are reporting results from the first high resolution X-ray
spectral observation of the RB, using the \axaf\ High Energy Transmission
Grating Spectrometer (HETGS).  A description of
the HETGS and its performance are given by Canizares et al.\
(in preparation).
Our objective was to search for specral emission and absorption
features that might distinguish the bursts from the persistent
emission.   The type II bursts, due to their E-$\Delta$t character,
are believed to be emission from the surface of the NS.  Discovery of
any emission or absorption features might then provide a measure of
the gravitational redshift at the NS photosphere.   

\section{Observations and Data Reduction}

\begin{figure}
\plotone{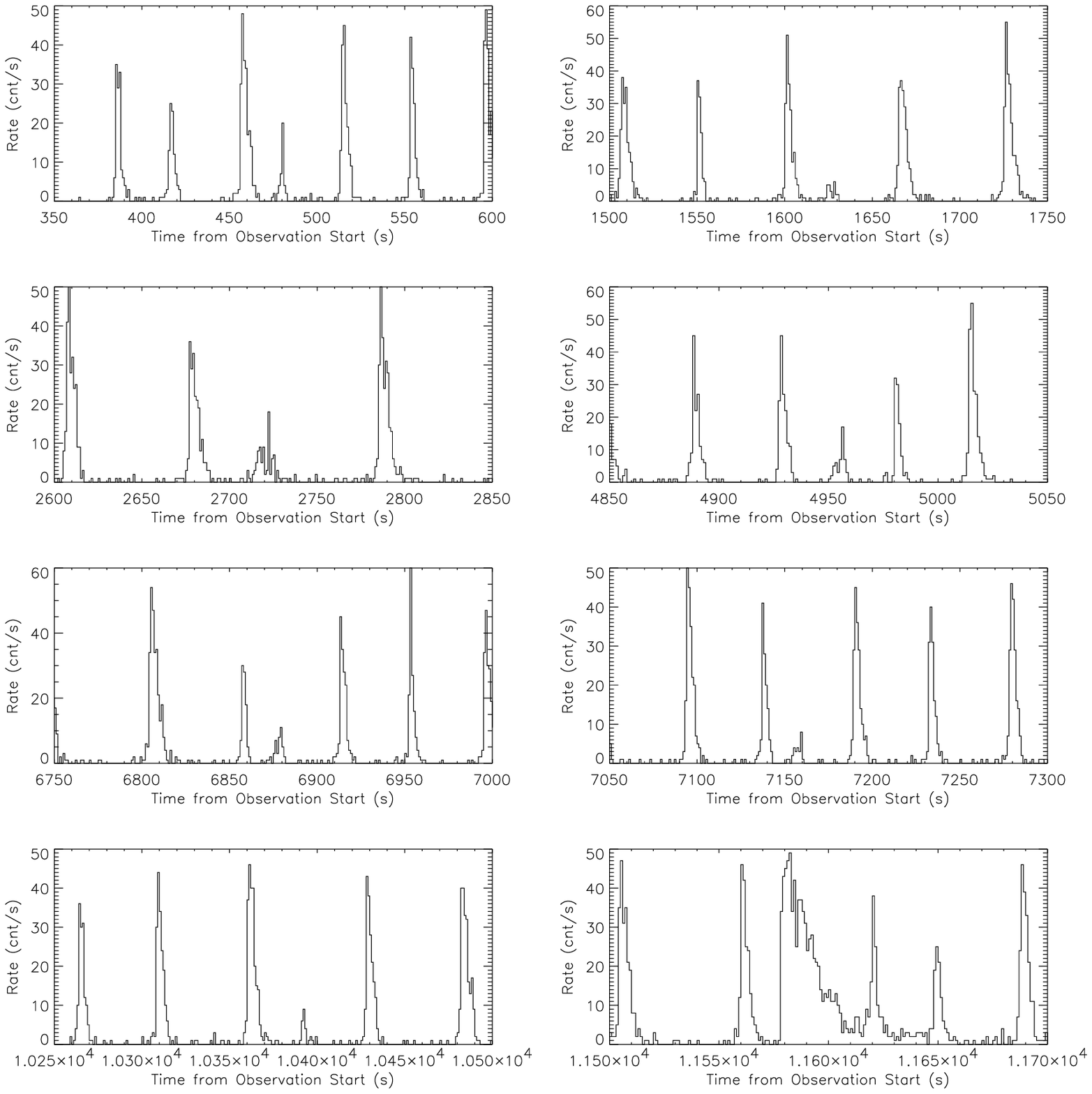}
\caption{Segments of the light curve from zeroth order.  The
times are offset to the beginning of the observation.  Most of
the bursts are the normal type II bursts from the Rapid Burster.
Near the midpoint of the time interval shown in the first seven panels 
there is a a ``mini-burst'' where the peak count rate is 5-40\%
of the average of the normal type II bursts.  The bottom right panel shows
the type I burst; the tail is so long that two type II bursts are
observed during the decay of the type I burst.
\label{fig:lightcurve} }
\end{figure}

The RB was observed with the \axaf\ HETGS on 24-25 October 1999
(MJD = JD - 2400000.5 = 51476.2-6.6).
The Advanced CCD Imaging Spectrometer (ACIS-S) detector was used
for readout and was operated in the so-called ``continuous clocking''
(CC) mode.  This mode is mentioned in the \axaf\ Proposer's
Observatory Guide (p. 101).
In this mode, the charge is shifted row by row to the frame
storage buffer at 2.84996 ms intervals so the X-ray events
are timed to an accuracy of about 3 ms by locating events within
the storage buffer, assuming that the event is associated
with the target.  The total exposure time
was 29562.7 s, during which a total of 702 bursts were
detected.  A few samples of the light curve from zeroth
order are shown in Fig.\ref{fig:lightcurve}, which shows
a type I burst about 11580 s into the observation.
The type I burst was fairly typical of those discovered by
\citet{hml78}, having a similar peak flux to that of the type II
bursts but with a longer decay time constant.

The median peak burst
count rate was about 45 count s$^{-1}$ but the average
burst count rate is about a factor of two smaller due to
the nearly triangular shapes of the burst profiles.
Average burst spectra were determined by selecting
events where the count rate over a 1 s interval exceeded
5 count s$^{-1}$, eliminating only 10\% of the total counts in type II
bursts while simultaneously reducing the background 
(see \ref{sec:spectra}) by more than a factor of three.
A ``peak'' burst spectrum was also derived by restricting
the selection to intervals where the count rate exceeded 30 count s$^{-1}$,
which gave half as many counts as the average spectrum.
After removing 82 s containing the type I burst,
the exposure time was 4326 s in the average spectrum and 1349 s
in the peak spectrum.
The average count rate was 23.4 count s$^{-1}$ for the average
spectrum and 39.7 count s$^{-1}$ in the peak spectrum.
In addition to the type I burst there were 18 ``mini-bursts'' where the
peak flux is 10-40\% of the peak flux from the remaining bursts which
are mostly included in the average spectrum but entirely excluded
from the peak spectrum.
The first seven of these mini-bursts are shown in Fig.\ref{fig:lightcurve}.

The CC mode introduces complications into the spectral processing.
Imaging information along
the readout direction is lost in favor of improved event timing.
In the timed exposure (TE) mode, the imaging information is used
to select the HEG and MEG data for spectral extraction and for
the selection of background.  In CC mode, only the ACIS pulse
height data are used.  The MEG and HEG first orders are well
separated in energy when at any specific dispersion distance
although there is the ambiguity that an event dispersed into 2nd
order by the MEG gratings will have the same energy as an HEG
photon and dispersed by the same distance.  In practice, however,
the efficiency of the MEG in 2nd order is a few percent of the
HEG efficiency in the 1-10 keV range, so this ambiguity is not
an issue.

Background selection for spectral analysis is not simple,
so this instrument mode is best suited
for bright targets such as the RB during bursts.  We cannot
define the background spatially so we instead select regions
of pulse height as a function of dispersion distance with the
same width, $\Delta E$, as used for selecting the source but
displaced in energy (see step 10 of the spectral processing).
Due to the difficulty in processing and analyzing the CC mode data,
we took great care to check our methods using data from other
sources observed with this mode, such as the Crab pulsar
(observation ID 170) and Cyg X-1 (observation ID 1511).

\subsection{Imaging}

As stated, there is limited spatial information; we
obtain $\sim$1\arcsec\ resolution along one
dimension of the detector array.  In this dimension,
when the spacecraft dither and the thermal drifts are removed, the result
is a very narrow image of the HETGS zeroth order.  It is well fit with a
Gaussian of 0.77\arcsec\ FWHM, which is consistent with the superb imaging
of the \axaf\ mirror assembly if the image is integrated along one
dimension.\footnote{Although the one dimensional response function is
not well documented, one may examine results posted on the web by
Dan Dewey:
{\tt http://space.mit.edu/HETG/technotes/zo\_1d/focus\_trends.010227.html}.}
In order to properly remove the telescope dither, however, we required
advancing the times assigned to the events by about 5 s.
Other instruments
on \axaf\ do not require this timing offset, so we conclude that the
offset is an unaccounted fixed timing offset.  It was also found to be
required for dither correction in all other CC mode observations, including
an observation of the Crab pulsar and the pulsar B0540-69
\citep{kaaret00}.\footnote{Note
that \cite{kaaret00} misstated the sign of the correction.
A constant offset of about 5 s must be {\em subtracted} from the times
of events in order to match the time of the aspect system.  Of five CC
mode observations we have processed, the sign and magnitude
of the offset is the same.  We estimate that the offset time is accurate
to about 0.5s.}

\subsection{Spectra}
\label{sec:spectra}

The spectral data were reduced starting from level 1 data provided
by the Chandra X-ray Center (CXC) using
IDL custom processing scripts; the method is quite similar to
processing data from timed exposure (TE) mode but has two wrinkles
that make it more difficult.  This procedure was successfully
verified using CC mode data from Cyg X-1 and is more thoroughly
described by Marshall, et al.\ (in preparation)
Briefly, the procedure was to:
1) restrict the event list to the nominal grade set (0,2,3,4, and 6),
2) remove the event streaks in the S4 chip (which are not
related to the readout streak and appear in CC and TE mode
data alike),
3) compute a projected dispersion distance ($x_p$, in mm) by
correcting for telescope dither along the dispersion direction
after applying the timing offset determined using the
dither pattern of the zeroth order events along the readout direction
(as described in the last section),
4) determine the location of zeroth order, $x_{p0}$, by
fitting a Gaussian to the one dimensional profile,
5) estimate the event wavelength $m\lambda = P (x_p-x_{p0}) /
(R \cos \theta )$
(where $P$ is the grating period, $m$ is the dispersion order, $R$
is the Rowland distance in mm, and $\theta$ is the clocking angle of the
MEG or HEG),
6) correct event energies for detector node-to-node gain variations,
7) select $\pm 1$ order events in order to minimize background
using $| E_{ACIS} * m \lambda / (h c) | - 1 < \Delta $ where $E_{ACIS}$
is the event energy inferred from the ACIS pulse height, and $\Delta$
is a function of $\lambda$, ranging from 0.025 at 1\AA\ to 0.15-0.20
for $\lambda > 6$ \AA,
8) eliminate events
in bad columns where the counts in an histogram deviate by 5$\sigma$
from a 50 pixel running median,\footnote{There were many
bad columns, especially on the S2 and S4 chips.  The cause may
be a poor on-board bias map calculation (Peter Ford, private
communication).  These mostly
affect the dispersed spectrum below 1 keV so they have a
negligible effect on our spectral results.}
10) select background events by applying a $E_{ACIS}$ criterion similar to
that applied to select source events but accepting events in the
range $\Delta$ to $2 \Delta$,
11) bin MEG (HEG) events at 0.01\AA\ (0.005\AA),
12) eliminate\ data affected by detector gaps, and
13) generate and apply an instrument effective area based on the
pre-flight calibration data and a set of correction polynomials derived
from the CC mode observation of Cyg X-1 (Marshall et al., in preparation).

\begin{figure}
\plotone{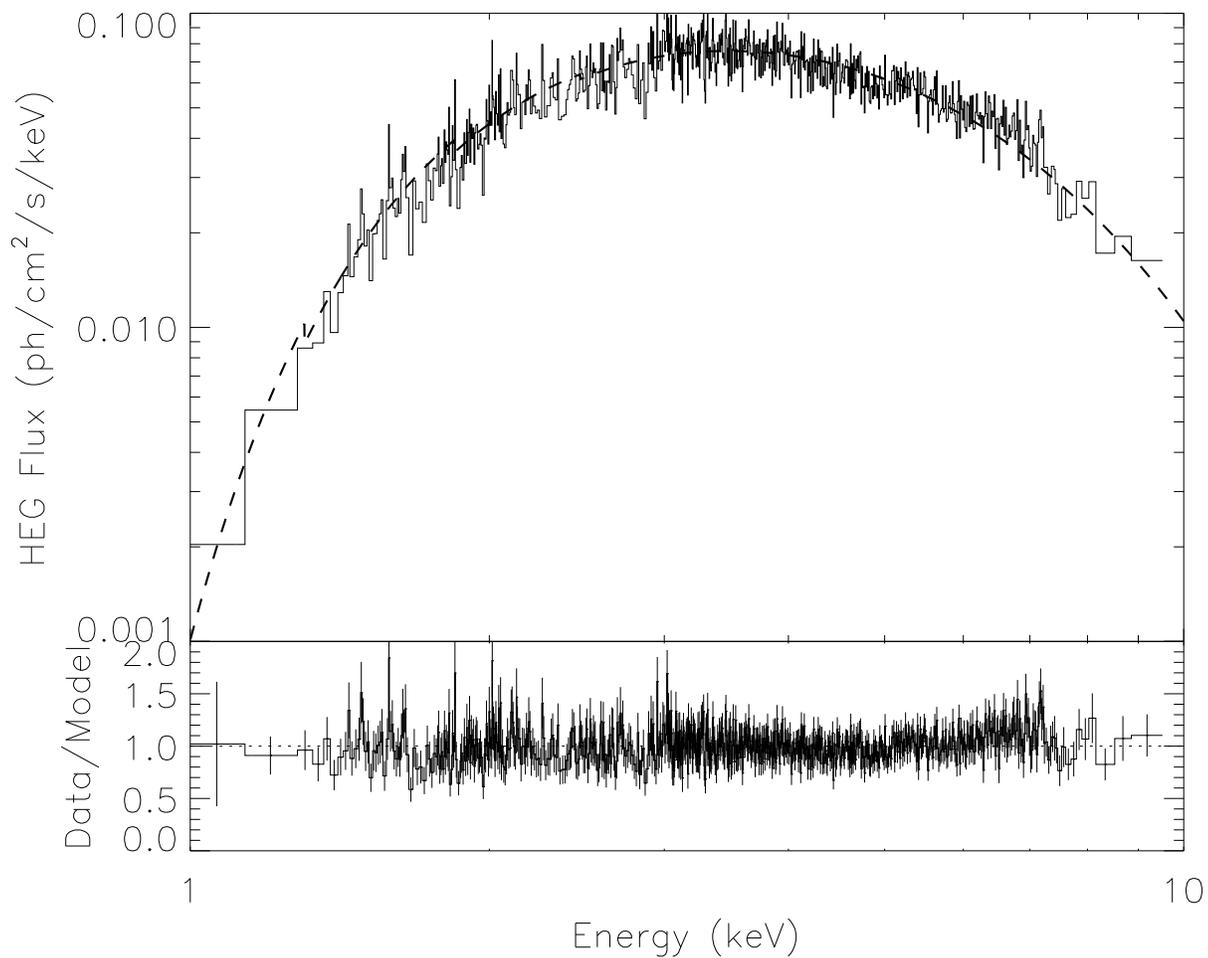}
\caption{Average burst spectrum from the high energy
gratings (HEG)s on the {\em Chandra}
HETGS.  The data have been adaptively smoothed to give a nearly constant
signal/noise ratio.  A blackbody model has been fit to the data and overplotted
in the upper panel.  The data/model ratio is shown in the lower panel,
along with the 1$\sigma$ uncertainties.  There are no clear examples of
narrow emission or absorption features.
\label{fig:hegspec} }
\end{figure}

\begin{figure}
\plotone{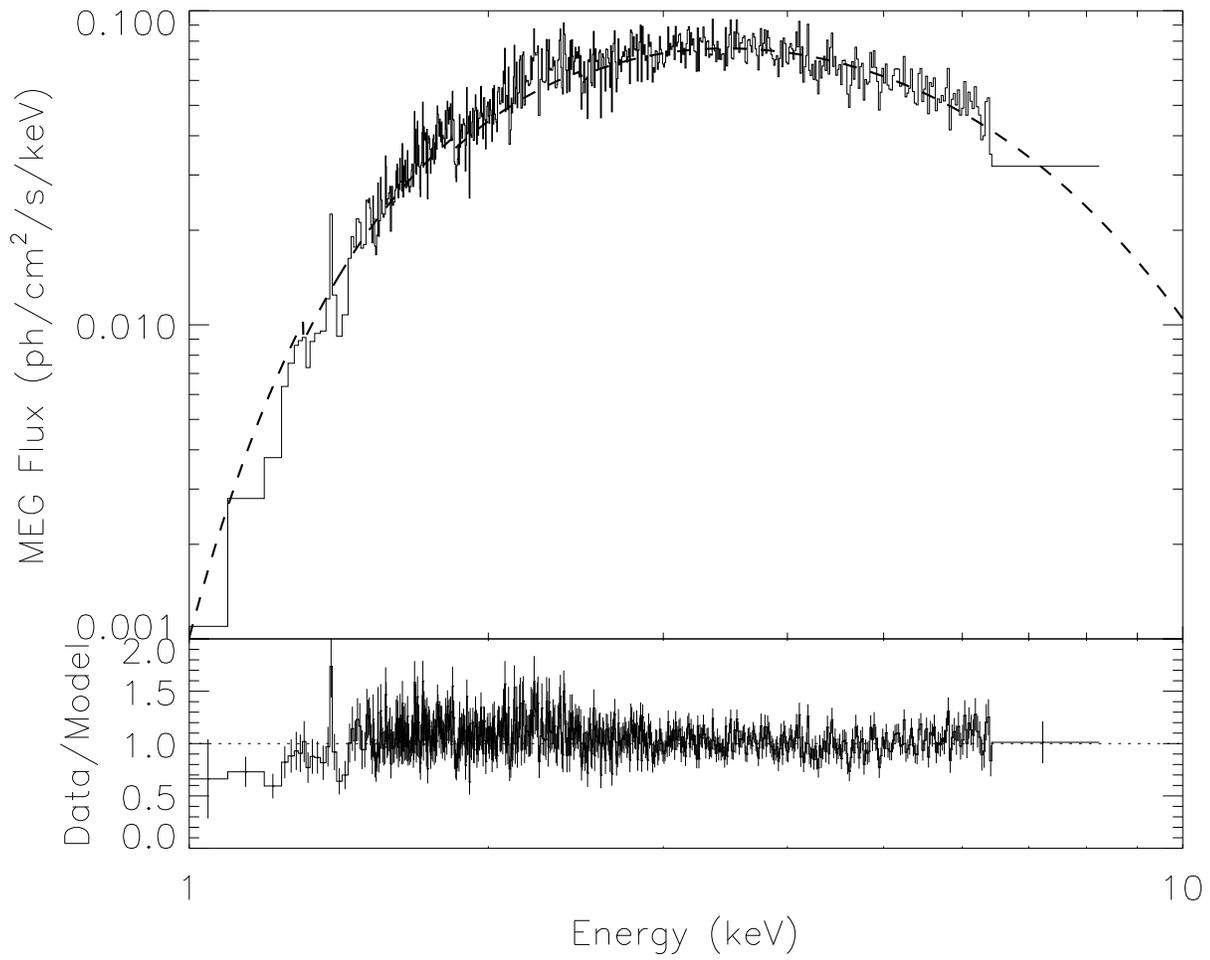}
\caption{Same as Fig.\ref{fig:hegspec} except for the medium energy gratings
(MEGs) on the {\em Chandra} HETGS.  Again, there are no clear examples of
narrow emission or absorption features.
\label{fig:megspec} }
\end{figure}

The time averaged spectra of the type II bursts
are shown in Figs.~\ref{fig:hegspec} and \ref{fig:megspec}.
The data have been adaptively rebinned to give a nearly constant
signal/noise ratio better than 10 except in the bin at the lowest
energy.  A blackbody model has been fit to the data and overplotted
in the upper panel.  A 5\% systematic error was added to the uncertainties
in quadrature so that the $\chi^2$ came out to 1.00.  This model
of the uncertainties is not precisely correct because there are
correlations between the systematic error in adjacent bins.
One systematic error is due to the difference between the
background spectra derived from events with
$E_{ACIS}$ larger than $(h c) / (m \lambda)$ compared to
those events where $E_{ACIS}$ is smaller than $(h c) / (m \lambda)$.
The background is less than 10\% of the signal for $1.5 < E < 8.0$ keV.
The systematic errors are small in this energy band so that their true
dependence on energy will not have a significant effect on our
results.

The burst peak spectra are well fit by a blackbody with
$kT = 1.590 \pm 0.016$ keV with a radius of
$8.88 \pm 0.14 $ km for a distance of 8.6 kpc, and
the interstellar column density, $N_H$, was measured to
be $1.81 \pm 0.03 \times 10^{22}$ cm$^{-2}$.  The fit to the
burst average spectra gave similar values for the
temperature and $N_H$, $1.577 \pm 0.012$ keV and
$1.74 \pm 0.03 \times 10^{22}$ cm$^{-2}$, but the fitted
radius was significantly smaller, $6.54 \pm 0.08$ km.
The uncertainties determined in the fit are indicative of
the high counting statistics ($> 10^5$ counts) but
would increase as one takes systematic errors into
account; we estimate that the
uncertainties should be regarded as good to a factor of 2.
In principle, we would be able to verify the fitted value of
$N_H$ by measuring the ionization edges due to Mg and Si at 1.30 and
1.84 keV, respectively.  With large $N_H$ and relatively low
source flux below the spectral peak of the blackbody emission,
the flux is insufficient to measure the optical depths at the
edges to better than a factor of 2, so we cannot obtain a useful
independent constraint on $N_H$.

\begin{figure}
\plotone{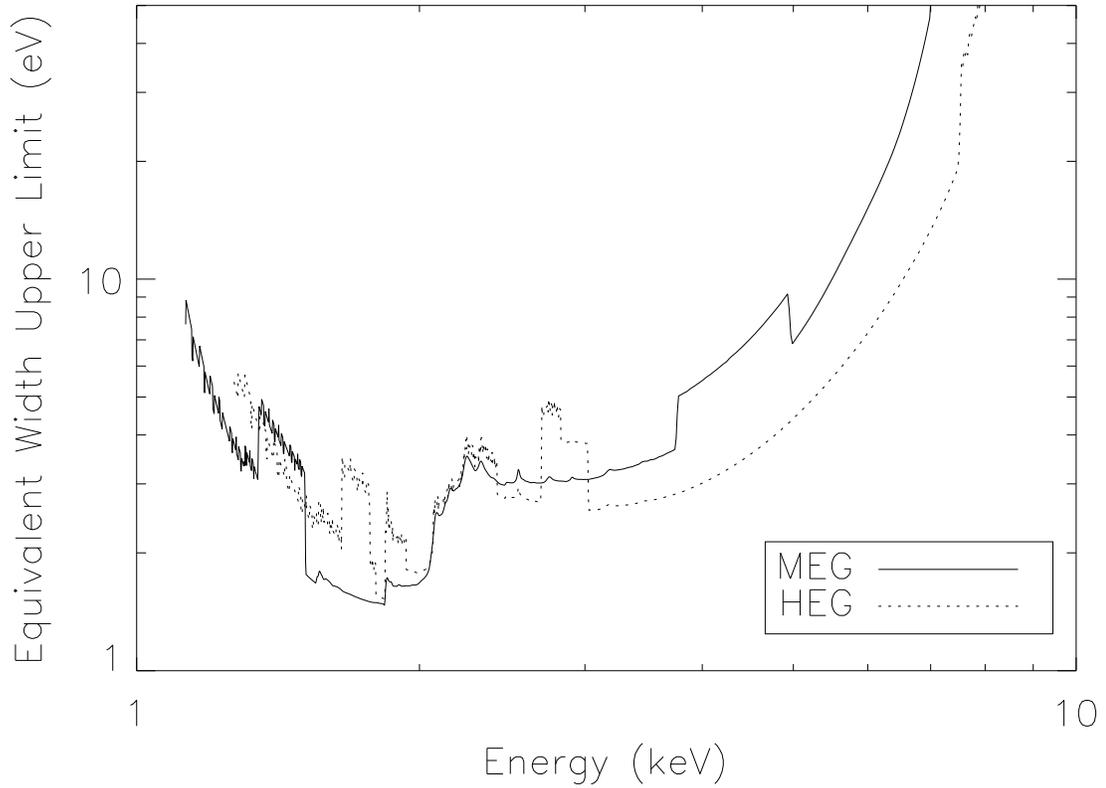}
\caption{Estimates of the 3$\sigma$ limits that can
be placed on any emission lines whose FWHMs are
comparable to the instrument resolution.  The computation
uses the model fitted to the time-averaged
HEG and MEG data and the effective areas.
We assume that candidate features are only 2 bins wide.
The curves are rather smooth, except
for locations of chip gaps, so one may derive limits on broad features using
these curves until the scale of the feature becomes comparable to that of the
instrument calibration uncertainties.  Limits on absorption
features are identical
when there are many counts but are systematically larger
at the high and low ends of the spectrum where there are less
than 25 counts per bin.
\label{fig:equivwidth} }
\end{figure}

We searched the time-averaged spectra for narrow absorption and emission
features and found several candidates.  All candidates were
found to be spurious, however, by comparing the HEG and MEG
spectra and the $+1$ orders to the $-1$ orders.  The burst
peak spectra also showed no features.
In Fig.~\ref{fig:equivwidth}, we show
estimates of the 3$\sigma$ limits that can be placed on
any emission lines whose widths (FWHMs)
are comparable to the instrument resolution.
The computation uses the model fitted to the time-averaged
HEG and MEG data and the effective areas.
We assume that candidate features are only
2 bins wide.  The curves are rather smooth, except
for locations of chip gaps, so one may derive limits on broader features using
these curves until the scale of the feature becomes comparable to that of the
instrument calibration uncertainties.
Limits on absorption lines are identical
when there are a lot of counts but are systematically larger.
The 3 $\sigma$ upper limits to
the equivalent widths of any features are
$<$ 10 eV in the 1.1-7.0 keV band and as small as
1.5 eV near 1.7 keV.

\section{Discussion}

We find that the color temperature at the peak
of the type II bursts is the same
as the average color temperature to within the very
small statistical uncertainties even though the count rate
drops by a factor of two from the average of the burst
peaks to the average of all type II bursts.
This finding confirms
a result that is well established for significantly
longer bursts -- that the color temperature is
not observed to vary significantly during the bursts \citep{marshall79}.
Thus, the profile of a type II burst seems to result
from variation of the size of the emission region
with a maximum of about 9 km.
If all emission takes place on the surface of
a neutron star, then any
smaller apparent radius must result from nonuniformities
over the stellar surface; in the simplest geometry,
the emission is assumed to emanate
from a spot on the surface.  During the fainter parts of
normal type II bursts and near the peaks of minibursts
the observed intensity drops to 10\% of the peak so
the apparent fraction of the surface, $f_a$, is about 0.1.
\citet{marshall82} computed $f_a$ of an emission
region spot covering a true fraction $f$ of the surface
for various values of the ratio of the stellar radius
to its Schwarzschild radius.
For a 1.4 $M_{\sun}$ neutron star with a 7 km radius
which is uniformly radiating across its surface,
the apparent radius (measured at infinity) is 11 km, so
we would infer that the type II emission region covers between
5 and 20\% of the surface when faint and an average of 50-85\%
at burst peak.  The endpoints
of the allowed range depend on whether the spot is
centered on the nearest or farthest points of the surface.
Gravitational bending of light trajectories ensures that all
bursts can be detected as long as at least 5-10\% of the
surface radiates, for relatively compact neutron stars.

The main objective of this observation was to determine
if there would be emission lines or absorption
features that could be used to determine the gravitational
redshift of the emission region.
At the estimated temperature of the emission region, Fe
will be highly ionized, so a significant fraction
of the Fe will be He-like Fe {\sc XXV},
for which the strongest resonance line is at 6.70 keV.
If the radiating region is spherically symmetric, then the
emission region is a shell at a radius of 7 km and the
Fe {\sc XXV} line will be gravitationally redshifted
to about 4.3 keV, where our 3$\sigma$ upper limit to absorption
features is 3 eV.  The instrumental resolution is 0.01\AA,
which is 15 eV at this energy.
Pressure broadening of the resonance line
due to the Stark effect in a highly ionized, hydrogen-dominated
plasma is estimated to be about 4 eV in the rest frame
of the gas while fine-structure splitting is
of order 21 eV \citep{paerels97}.  There is a
30\% narrowing to the observed frame due
to the gravitational redshift so the line FWHM is expected
to be comparable to the instrumental resolution.
The thermal line FWHM is smaller,
$\sim 1$ eV.  Thus, the upper limit on the column density of
Fe {\sc XXV} is about $8 \times 10^{15}$ cm$^{-2}$.
For a simple model atmosphere dominated by electron scattering,
the column density of Fe to one optical depth is of order
$X_{Fe} / \sigma_T = 5 \times 10^{19}$ cm$^{-2}$, where $X_{Fe}$
is the abundance of Fe and $\sigma_T$ is the Thomson
cross section.  The discrepancy of $\sim 10^4$ leads us
to suggest that Compton scattering dominates the spectrum,
destroying possible line features.

At the measured color temperature, Comptonization is expected to
shift low energy photons to higher energies.  The color
temperature is systematically larger than the effective
temperature of the plasma.  The magnitude of this bias
is not well determined but can be a factor of 1.5 (cf. LVT).
A decrease of the temperature requires a compensating increase
of the inferred radius of the emission region, yielding a
radius of approximately 20 km.
This radius is somewhat larger than observed in other burst
sources but still feasible from theoretical standpoint.

\acknowledgments

This work has been supported in part under NASA grant NAG5-3239, and
NASA contracts NAS8-38249, NAS8-39073, and SAO SV1-61010.

\clearpage

\clearpage

\end{document}